\documentstyle{article}

\catcode`@=11
\newif\ifIncludeFigs\IncludeFigsfalse

\IncludeFigstrue

\ifIncludeFigs
      \@input{epsf} %
      \ifx\epsfbox\undefined
         \EPSFfalse
	 \IncludeFigsfalse
      \else
      \fi
\else
\fi

\ifx\endfloat\notincluded\else
  \nomarkersintext
\fi

\catcode`@=11

\ifx\@ove@rcfont\undefined
  \ifx\citen\undefined
     \def\citen#1{\begingroup \def\@cite##1##2{{##1}}%
	\@citex[]{#1}\endgroup}
  \fi
\else
  \def\@cite#1{$\@ove@rcfont\m@th^{[{#1}]}$}
\fi

\def\caption{\@dblarg\aux@caption}

\def\aux@caption[#1]#2{
   \parindent 20pt \par
      {\refstepcounter\@captype \@caption{\@captype}[#1]{#2}}}

\long\def\@caption#1[#2]#3{\par\addcontentsline{\csname
  ext@#1\endcsname}{#1}{\protect\numberline{\csname
  the#1\endcsname}{\ignorespaces #2}}\begingroup
    \@parboxrestore
    \small\sl
    \@makecaption{\csname fnum@#1\endcsname}{\ignorespaces #3}\par
  \endgroup}

\def\maketitle{
 \begingroup
 \def\thefootnote{\fnsymbol{footnote}}
 \def\@makefnmark{\hbox
 to 0pt{$^{\@thefnmark}$\hss}}
 \if@twocolumn
 \twocolumn[\@maketitle]
 \else 
 \global\@topnum\z@ \@maketitle \fi
 \thispagestyle{empty}
 \setcounter{page}{0}
 \@thanks
 \endgroup
 \setcounter{footnote}{0}
 \let\maketitle\relax
 \let\@maketitle\relax
 \gdef\@thanks{}\gdef\@author{}\gdef\@title{}\let\thanks\relax}

\def\paperid#1{\gdef\@paperid{#1}}

\def\@maketitle{

 \@makepub
 \vskip 4em \begin{center}
 { \Large \bf \@title \par}
 \vskip 1.5em {\large \lineskip   .5em
 \@authoraddress
 }
 \end{center}
 \par
 \vskip 1.5em
}

\def\@makepub{{
  \centering
  \makebox[\textwidth]{
    \parbox[t]{0.25\textwidth}{\begin{flushleft}%
      {\small\@pubdate}\end{flushleft}}
    \hfil
    \parbox[t]{0.5\textwidth}{\begin{center}%
      {\small \@publabel}\end{center}}
    \hfil
   \parbox[t]{0.25\textwidth}{\begin{flushright}{\small
    \@pubnumber}\end{flushright}}
  }
}}

\gdef\@publabel{\hfil}
\gdef\@pubdate{Jan 1, 1999}
\gdef\@pubnumber{EFI-??-??}

\long\def\pubdate#1{\gdef\@pubdate{#1}}
\long\def\pubnumber#1{\gdef\@pubnumber{#1}}
\long\def\publabel#1{\gdef\@publabel{#1}}

\else
  \font \crest=uofc-shield
  \gdef\@publabel{\crest C}
\fi
\def\abstract{\if@twocolumn
\section*{ABSTRACT}
\else \small
\begin{center}
{
ABSTRACT\vspace{-.5em}\vspace{0pt}}
\end{center}
\quotation
\fi%
}

\def\endabstract{\if@twocolumn\else\endquotation\fi}

\def\pacs#1{\par %
  \bgroup
  \hsize\columnwidth \parindent0pt
  \if@twocolumn\else\leftskip=0.10753\textwidth \rightskip\leftskip\fi
 \ifdim\prevdepth=-1000pt \prevdepth0pt\fi
 \dimen0=-\prevdepth \advance\dimen0 by20pt\nointerlineskip
  \vbox to28pt{\small\vrule height\dimen0 width0pt\relax%
	PACS: #1\vfill}
  \egroup
  \if@twocolumn\vskip1pc\fi
  \newpage
}

\gdef\@author{Nobody}
\gdef\@authoraddress{}

\def\@makeauthor{
  {\def\and{\smallskip {\normalsize \rm and\smallskip}}
  {\zerospfalse \centering \large \@author}
  }
}

\def\author#1{\expandafter\def\expandafter\@authoraddress\expandafter
  {\@authoraddress %
  {
  \dimen0=-\prevdepth \advance\dimen0 by23pt
  \nointerlineskip
  \rm\centering
  \vrule height\dimen0 width0pt\relax\ignorespaces#1%
      \\[\baselineskip] 
  }%
  }%
}

\def\address#1{\expandafter\def\expandafter\@authoraddress\expandafter
  {\@authoraddress{\small\it\centering \baselineskip 1.3\baselineskip
\ignorespaces#1 \par
  }}
}

\def\thebibliography#1{\newpage
\section*{References\markboth{REFERENCES}{REFERENCES}}
\addcontentsline{toc}{section}{References}\labelsep1.0em\list
  {\arabic{enumi}.}{\settowidth\labelwidth{#1.}%
  \leftmargin\labelwidth
    \advance\leftmargin\labelsep\usecounter{enumi}}}

\let\acknowledgements=\acknowledgement

\def\msubsection{\@startsection{subsection}{2}{0.25em}%
  {4.5ex plus 1ex minus .2ex}{1.0ex plus .2ex}{\normalsize\bf}}

\def\appendix{\par
  \setcounter{section}{0}
  \setcounter{subsection}{0}
  \setcounter{equation}{0}
  \def\theequation{\thesection.\arabic{equation}}
  \def\thesection{\Alph{section}}}

\advance \topmargin -3\baselineskip
\headheight=\baselineskip \headsep=2\baselineskip
\def\twoup{
        \mytwocolumn
	\sloppy\flushbottom\parindent 2em
        \leftmargini 2em\leftmarginv .5em\leftmarginvi .5em
        \oddsidemargin -.5in    \evensidemargin 0in
        \columnsep .4in \footheight 0pt
        \textwidth 10in \topmargin  -.4in
        \headheight 0pt \topskip 0in
        \textheight 6.9in \footskip 30pt
        \def\@oddfoot{\hfil\thepage\hfil\addtocounter{page}{1}
                \hspace{\columnsep}\hfil\thepage\hfil}
        \let\@evenfoot\@oddfoot \def\@oddhead{} \def\@evenhead{}
}

\def\mytwocolumn{
   \global\columnwidth\textwidth
   \global\advance\columnwidth -\columnsep \global\divide\columnwidth\tw@
   \global\hsize\columnwidth \global\linewidth\columnwidth
   \global\@twocolumntrue \global\@firstcolumntrue
   \@dblfloatplacement\@ifnextchar[{\@topnewpage}{} 
}

\newdimen\slashraise \slashraise=0.33pt
\mathchardef\fslash="0236

\def\slash@char#1#2{%
   \setbox0=\hbox{$\m@th#2$}
   \dimen0=\wd0                                 
   \dimen2=-\dp0 \advance\dimen2 by \slashraise
   \setbox1=\hbox{$\m@th#1\mkern-13mu\fslash$}
	 \dimen1=\wd1               
   \ifdim\dimen0>\dimen1                        
      \rlap{\hbox to \dimen0{\hss\raise\dimen2\box1\hss}}%
      #2                          		
   \else                                        
      \rlap{\hbox to \dimen1{\hss\box0\hss}}    
      \raise\dimen2\box1                              
   \fi}                                         %

\def\slashchar#1{\mathpalette\slash@char#1}
\let\slsh=\slashchar

\let\mdef\def

\def\lesssim{\mathrel{\mathpalette\vereq<}}
\def\vereq#1#2{\lower3pt\vbox{\baselineskip1.5pt \lineskip1.5pt
\ialign{$\m@th#1\hfill##\hfil$\crcr#2\crcr\sim\crcr}}}

\mdef\bar{\overline}
\mdef\eqb{\begin{equation}}
\mdef\eqe{\end{equation}}

\mdef\MeV{{\rm \,MeV}}
\mdef\GeV{{\rm \,GeV}}

\mdef\ie{{\it i.e.}}
\mdef\etal{{\it et al.}}

\let\goesto\rightarrow

\mdef\Or#1{O(#1)} 
\mdef\Order{O}

\mdef\BRA#1{\left\langle #1\right|}
\mdef\bra#1{\langle #1 |}
\mdef\KET#1{\left| #1\right\rangle}
\mdef\ket#1{| #1\rangle}
\mdef\VEV#1{\left\langle #1\right\rangle}
\mdef\vev#1{\langle 0 | #1 | 0\rangle}

\def\tr{\mathop{\rm tr}\nolimits}
\def\str{\mathop{\rm str}\nolimits}
\def\sdet{\mathop{\rm sdet}\nolimits}
\def\det{\mathop{\rm det}\nolimits}

\def\trd{\mathop{\rm tr}_D\nolimits}

\mdef\lr{\leftrightarrow}

\mdef\scr#1{{\cal #1}}

\mdef\chisim{$SU(3)_L \times SU(3)_R$}
\mdef\im{{\rm i}}
\mdef\pintegral{\int\!{d^D\,p\over(2\pi)^D}}

\mdef\L#1{\,\mu_{#1}}
\mdef\Lq{\,\mu_q}
\mdef\LM{\L{d}}

\mdef\onpif#1{{#1 \over 16 \pi^2 f^2}}

\mdef\qt{{\tilde q}}
\mdef\qbar{{\overline q}}
\mdef\qtbar{{\overline\qt}}
\mdef\phit{{\tilde\phi}}

\mdef\Ht{{\tilde H}}
\mdef\Bt{{\tilde B}}
\mdef\etat{{\tilde \eta}}
\mdef\epshat{\hat\epsilon}

\publabel{ \ 
	}
\pubdate{July, 1994 \\ Revised November, 1994}
\pubnumber{IFT-94-09 \\ hep-ph/9411433
 }

\begin{document}
\begin{titlepage}
\title{Quenched Chiral Perturbation Theory for Heavy-Light Mesons}

\author{Michael J.~Booth
}
\address{%
  Institute for Fundamental Theory and Department of Physics\\
              University of Florida, Gainesville, Florida 32611 \\
		 {\tt booth@phys.ufl.edu}
}
\maketitle
\begin{abstract}
Quenched chiral perturbation theory is extended
to include heavy-light mesons.  Non-analytic corrections to the decay
constants, Isgur-Wise function and masses and mixing of heavy mesons
are then computed.
The results are used to estimate the error due to quenching in lattice
computations of these quantities.
For reasonable choices of parameters,
it is found that quenching has a strong effect on $f_{B_s}/f_B$,
reducing it by as much as $28\%$.  The errors are essentially
negligible for the Isgur-Wise function and the mixing parameter.

\end{abstract}
\centerline{(To Appear in Physical Review D)}

\end{titlepage}
\section{Introduction}

Lattice simulations of hadron properties
have made great progress in recent years and there is hope
that they will soon
yield accurate ``measurements'' of quantities that are difficult or
impossible to access experimentally, such as the kaon
mixing parameter $B_K$ and the $B$ and $D$ decay
constants $f_B$ and $f_D$, play an important role in the
phenomenology of the standard model.
This progress
has come not only through improvements in computer speed and
algorithms but also through better understanding of errors.
One systematic error
present in most calculations is that
arising from the use of the quenched (or valence) approximation, in
which disconnected fermions loops are neglected.
For heavy quarks, \ie\ those with masses well
above the QCD scale, such as the $b$ and $c$, the decoupling theorem
ensures that quark loops can be accounted for by suitable adjustments
of the coupling constants.  But for lighter quarks,
with masses below the
QCD scale,  it is expected that
quenching will change not only the short-distance but also
the long-distance properties of the theory;
these latter changes are much more difficult to quantify.

A straightforward way to study this error is to perform simulations
with dynamical fermions and compare the results to similar
calculations in the quenched approximation.  However, this is still a
complicated undertaking because calculations with dynamical fermions
are performed with larger lattice spacings and heavier quarks.  This
of course increases the errors and makes it more difficult to
isolate the effect of quenching.
For example, the study of $f_B$ in Ref.~\citen{HEMCGC:fB} sees little
effect due to quenching, but the interpretation is difficult because
of the large lattice spacing used.

Another approach to understanding the error is to study how quenched
QCD differs from full QCD in the continuum.  That is, one compares the
quenched and full QCD predictions for a given quantity.  Then, to the
extent that lattice calculations reproduce the continuum theory, the
difference between the two predictions gives an indication of the
error due to quenching.  This analytic approach was initiated by
Morel\cite{Morel}, who studied how chiral logarithms differ in the two
theories.  It was extended by Sharpe\cite{Sharpe:BK}, who developed a
diagrammatic analysis, and later Bernard and Golterman\cite{BG}
formulated quenched chiral perturbation theory to discuss these
logarithms in a systematic way.  Corrections to light meson decay
constants and masses, $B_K$ and recently baryon masses\cite{LabSharpe}
have been studied using these techniques.

In this paper I will extend quenched chiral perturbation theory
to include heavy-light mesons.  This enables one to study the effect
of quenching on lattice studies of these mesons.
The paper is organized as follows.
In section 2, I
review the combination of chiral and heavy quark symmetries.  I
continue the review by showing how chiral perturbation theory
can be formulated for
the quenched approximation to QCD.
Finally I show how to extend this to include heavy mesons.
In section 3, I compute loop corrections to the heavy meson
decay constants and mass splittings, the mixing
parameter $B_B$ and the Isgur-Wise function $\xi$.
In section 4, following a discussion of the parameters of
the theory, the results are investigated numerically.
In section 5 I conclude and comment on possibilities for
future study.  An appendix collects results for
the renormalized couplings.

\section{Quenched Chiral Perturbation Theory and the Inclusion of
Heavy Mesons}

\subsection{Chiral Theories}
%

To lowest order in the chiral expansion, the self-interactions of
the light mesons are described by the Lagrangian
\eqb
\label{eqn:ChL}
\scr{L} = {f^2\over 8} \left[
        \tr(\partial_\mu\Sigma\partial^\mu\Sigma^\dagger)
       +2\mu_0\tr(M\Sigma + M\Sigma^\dagger)\right]
\eqe
with $\Sigma = \xi^2$ and
\eqb
\xi = e^{i\phi(x) / f},
\eqe
where
the light mesons are grouped into the usual matrix
\eqb
\phi =
\pmatrix{\frac 1{\sqrt 2}\pi^0 + \frac 1{\sqrt 6}\eta &
\pi^+ & K^+ \cr
\pi^- & -\frac 1{\sqrt 2} \pi^0 + \frac 1{\sqrt 6}\eta &
K^0 \cr
K^- & {\bar K}^0 & -\sqrt{\frac 2 3} \eta \cr}.
\eqe
The normalization is such that $f_\pi = 128 \MeV$.
Under $SU(3)_L \times SU(3)_R$, $\xi$ transforms as
\eqb
\xi \mapsto L \xi U^\dagger = U \xi R^\dagger.
\eqe
This equation implicitly defines $U$
as a function of $L$, $R$, and $\xi$.
The quark mass matrix%
\footnote{There should be no confusion when $M$ is also used to
   denote a generic heavy meson mass.}
$M$ is given by
\eqb
M = \pmatrix{m_u &&\cr & m_d &\cr && m_s \cr}.
\eqe
For purposes of determining the allowed form of the Lagrangian,
$M$ is given the ``spurion'' transformation rule
\eqb
M \mapsto L M R^\dagger,
\eqe
so it is convenient to define the quantities
\eqb
M_\pm = \frac 12 (\xi^\dagger M \xi^\dagger \pm \xi M \xi),
\eqe
which transform as
\eqb
M_\pm \mapsto U M_\pm U^\dagger.
\eqe

At leading order in the $1/M$ expansion,
strong interactions of $B$ and $B^*$ mesons are governed by the
chiral Lagrangian\cite{ChHQ}
\eqb
\label{eqn:ChHQL}
\scr{L} =
    -\trd\left[\overline H_a(v)\im v\cdot D_{ba} H_b(v)\right]%
    + g\,\trd\left[\bar H_a(v)H_b(v)\slsh{A}_{ba}\gamma_5\right]\,.
\eqe
The $B$ and $B^*$ fields are incorporated into
the $4 \times 4$ matrix $H_a$ which conveniently encodes the
heavy quark spin symmetry:
\begin{eqnarray}
H_a &=& {1\over2}(1+\slsh v)
	[\bar B^{*\mu}_a\gamma_\mu - \bar B_a\gamma_5],\\
\bar H_a &=&\gamma^0 H_a^\dagger \gamma^0\,.
\end{eqnarray}
Here $v^\mu$ is the four-velocity of the heavy meson,
the index $a$ runs over the light quark flavors, $u$, $d$, $s$ and
the subscript ``D'' indicates that the trace is taken only over
Dirac indices.
Henceforth I will
drop explicit reference to the heavy meson velocity.
Under \chisim, $H$ transforms as
\eqb
H \mapsto H U^\dagger.
\eqe
The light mesons
enter the heavy meson Lagrangian Eq.~(\ref{eqn:ChHQL}) through
the quantities:
\begin{eqnarray}
D_\mu &=& \partial_\mu + V_\mu, \nonumber \\
V_\mu &=& \frac 12\left(\xi \partial_\mu \xi^\dagger
+ \xi^\dagger \partial_\mu \xi\right), \\
A_\mu &=& \frac i2\left(\xi \partial_\mu \xi^\dagger
-\xi^\dagger \partial_\mu \xi\right) = - {1\over f}\partial_\mu \phi +
	O(\phi^3),
\end{eqnarray}
It follows from the definitions that under \chisim\
\eqb
V_\mu \mapsto U V_\mu U^\dagger + iU\partial_\mu U^\dagger, \qquad
A_\mu \mapsto U A_\mu U^\dagger,
\eqe
while the covariant derivative transforms as
\eqb
\label{eqn:covder}
D_\mu X \mapsto U D_\mu X U^\dagger.
\eqe
Finally, the left-handed current which mediates the
	decay $B\goesto l\nu$ is
represented by
\eqb
\label{eqn:current}
   J_a^\mu =  i\alpha \trd[\Gamma^\mu H_b\xi^\dagger_{ba}]
\eqe
where $\Gamma^\mu = \gamma^\mu L = \gamma^\mu (1 - \gamma_5)/2$.
At lowest order the decay constants are related
(in my normalization) by
$f_B$ = $\alpha/\sqrt{M_B}$,  $f_{B^*}$ = $\alpha \sqrt{M_B}$.

\subsection{Quenched QCD}
In the quenched approximation to QCD, the determinant which arises
in the functional integral when the
quark fields are integrated out, is omitted.
This can be implemented
in a formal way by introducing
for each quark $q_a$ a ``ghost'' partner $\qt_a$ with the same mass,
but bosonic statistics,
so that the ghost determinant cancels the quark
determinant\cite{Morel}.  The Lagrangian is then
\eqb
\label{eqn:qLag}
\scr{L}_{\rm quenched} =
   \sum_a\qbar_a(\slsh{D}+m_a)q_a
	+ \sum_a\qtbar_a(\slsh{D}+m_a)\qt_a.
\eqe

Classically,
when the masses vanish, the quenched Lagrangian (\ref{eqn:qLag}) is
invariant under the graded group $U(3|3)_L\times U(3|3)_R$,
but at the quantum level the full symmetry is broken by the anomaly%
\footnote{The broken $U(1)$ is that which acts
as $q \goesto e^{i \alpha \gamma_5} q$,
$\qt \goesto e^{-i \alpha \gamma_5} \qt$.}%
 to the semi-direct product\cite{BG}
$(SU(3|3)_L\times SU(3|3)_R){\bigcirc\kern -0.75em s\;}U(1)$.
Elements of the graded symmetry group are represented by supermatrices
(in block form)
\eqb
U=\left(\matrix{A&B\cr C&D\cr}\right),
\eqe
where $A$ and $D$ are matrices composed of even (commuting) elements
and $B$ and $C$ are composed of odd (anti-commuting) elements.  If we
assume that chiral symmetry breaks in the usual way, then the dynamics
of the remaining 18 Nambu-Goldstone bosons and the 18 Nambu-Goldstone
fermions can be described by an effective chiral Lagrangian, just as
for full QCD
\cite{BG,Sharpe:BK,BG2,Sharpe2,LabSharpe}.  The meson matrix is
extended to a supermatrix
\eqb
\Phi = \left(\matrix{\phi&\chi^\dagger\cr
\chi&\phit\cr}\right),
\eqe
where $\chi^\dagger \sim \qt \qbar$, $\chi \sim q \qtbar$ and
$\phit \sim \qt\qtbar$.  Note that $\chi$ and $\chi^\dagger$
are fermionic fields, while $\phi$ and $\phit$ are bosonic.
Group invariants are formed using the
super trace $str$ and super determinant $sdet$, defined as
\begin{eqnarray}
\str(U) &=& \tr(A)-\tr(D), \\
\sdet(U) &=& \exp(\str\log{(U)})=\det(A-B D^{-1} C)/\det(D).
\end{eqnarray}
The lowest order Lagrangian would then have the same form as
Eq.~(\ref{eqn:ChL})
above, with obvious notational changes.
%
But because the full symmetry group is broken by the anomaly,
extra terms are required
to describe the dynamics
of the anomalous field.  In full QCD, this anomalous
field is the $\eta'$,
and these extra terms can be neglected because the anomaly pushes
the mass of the $\eta'$ up beyond the chiral scale.  However, in
the quenched
theory, because of the absence of disconnected quark loops, this
decoupling does not occur: the super-$\eta'$ remains in the
theory and
the extra terms must be included.
To lowest order,
the complete Lagrangian is then
%
%
\eqb
\label{eqn:LBG}
\scr{L}_{BG} = {f^2\over 8} \left[
        \str(\partial_\mu\Sigma\partial^\mu\Sigma^\dagger)
        +4\mu_0\,\str(\scr{M}_{+}) \right]
    + \frac{\alpha_0}2 \partial_\mu\Phi_0\partial^\mu\Phi_0
    - \frac{m_{0}^2}2 \Phi_{0}^2
\eqe
with
\begin{eqnarray}
\Phi_0 & = &{1\over\sqrt3} \str \Phi =
	{ 1\over \sqrt2}(\eta' - \etat'), \\
\scr{M} &=& \left(\matrix{M&0\cr 0&M\cr}\right),
\end{eqnarray}
and $\scr{M}_{\pm}$ defined analogously to $M_{\pm}$.

The propagators that are derived from this Lagrangian are the ordinary
ones, except for the flavor-neutral mesons, for which the
non-decoupling
of $\Phi_0$ leads to a curious double-pole structure.  For these
mesons it is convenient to use a basis $U_i$, corresponding to
$u\bar u, d\bar d$ and so on, including the ghost quark counterparts.
Then the propagator takes the form
\eqb
G_{ij} = {\delta_{ij} \epsilon_i \over p^2 - M_i^2}
     + {(-\alpha_0 p^2 + m_0^2)/3 \over (p^2 - M_i^2)(p^2 - M_j^2)}
\eqe
where $\epsilon = (1,1,1,-1,-1,-1)$ and $M^2_i = 2\mu_0 m_i$.  It is
convenient to treat the second term in the propagator as a new vertex,
the so-called hairpin, with the rule that it can be inserted only once
on a given meson line.

Heavy mesons can be incorporated into this framework by adding to $H$
extra fields $\Bt$ and $\Bt^*$ derived from the heavy fields $B$ and
$B^*$ by replacing the light quark with a ghost quark.  It is
necessary to
include in the Lagrangian vertices which couple $\Phi_0$ to $H$.
Symmetry
requires that this coupling occur
through $\str(A_\mu)$, which no longer vanishes.
%
Including also
explicit $SU(3)$ breaking terms,
the Lagrangian is
\begin{eqnarray}
\label{eqn:FullL}
{\cal L}&=&
  -\trd\left[\bar H_a i v\cdot D_{ba} H_b\right]
  + g \trd\left[\overline
             H_a H_b \,\slsh{A}_{ba}\gamma_5\right]
  + \gamma \trd\left[\overline
             H_a H_a \gamma_\mu\gamma_5\right] \str(A^\mu)
  \nonumber \\
  &&\mbox{}+ 2\lambda_1 \trd\left[\overline H_a H_b \right] (\scr{M}_{+})_{ba}
    + k_1 \trd\left[
            \bar H_a i v\cdot D_{bc} H_b\right](\scr{M}_+)_{ca}
  \nonumber \\
    &&\mbox{}+  k_2 \trd\left[
            \bar H_a i v\cdot D_{ba} H_b\right]\str(\scr{M}_+).
\end{eqnarray}
The $B$ and $B^*$
propagators are
${i\over 2v\cdot k}$
and
${-i(g_{\mu\nu}-v_\mu v_\nu)\over 2 v\cdot k }$,
respectively; the ghost mesons have the same propagators as
their real counterparts.
To the same order, the current is given by
\begin{eqnarray}
\label{eqn:currentfull}
J_a^\mu &=&  i\alpha \trd[\Gamma^\mu H_b\xi^\dagger_{ba}]
+ i \alpha \kappa_1
      \trd[\Gamma^\mu H_c\xi^\dagger_{ba}] (\scr{M}_+)_{cb}
+ i \alpha \kappa_2
      \trd[\Gamma^\mu H_b\xi^\dagger_{ba}] \str(\scr{M}_+).
\end{eqnarray}
In the sequel, the terms proportional to $m_q$ will
be loosely referred to as counter-terms because they are required
to absorb the divergences encountered in loop calculations.
In addition, the presence of the additional mass scale $m_0$ means
there will new divergences (not found in the unquenched theory)
proportional to it.  For completeness, the divergent portions
of the counter-terms can be found in the appendix.

At this point let me pause to note a few peculiarities of the
theory just formulated.  First, while the symmetry
allows terms involving $\str(\scr{M_+})$, they do not contribute
at tree or one-loop level to any of the quantities I will consider
(although they do contribute in the unquenched theory).
Second, the loop structure of the quenched theory is rather odd.
Because the heavy mesons contain only one light quark and
there are no disconnected quark loops, none of the meson loops
involve any flavor changing vertices.
Consequently, the loop corrections
for a generic heavy meson $B_q$ containing the light quark $q$
will be a function of $M_q$ alone. The three-flavor theory is then
just three copies of a single-flavor theory.  This tends to heighten
the difference between the full and quenched theories because
not only are the virtual quarks lost, but the ``averaging'' effect
arising from the interaction with light mesons of different
mass is also lost.  This is in contrast with the situation for
light mesons in quenched QCD, where a kaon
has loop corrections involving $d\bar d$, $s\bar s$ and $d \bar s$
mesons, each (potentially) having a different mass.  It is the
cancelation between these
different meson loops, for example, which is behind
Sharpe's result\cite{Sharpe:BK} that $B_K$
is the same in the full and quenched
theories when $m_s = m_d$.

\section{Loop Corrections}
\subsection{Loop Integrals}

There are several loop integrals which will be encountered.
Two of these integrals are shown below.  The first
(here $D = 4 - 2\epsilon$, $1/\epshat =
 1/\epsilon + \log 4\pi - \gamma_{\rm E} + 1$).
\begin{equation}
\label{eqn:Logint}
    \im\pintegral {1\over p^2-m^2}=
	{-m^2\over16\pi^2}{1\over\epshat} +
	{1\over16\pi^2} m^2 \log(m^2/\mu^2)\,,
\end{equation}
arises from light meson tadpoles,
while the heavy-light loops require
\begin{eqnarray}
\label{eqn:Jint}
%
J^{\mu\nu}(m,\Delta)&=&
    \im\pintegral{p^\mu p^\nu\over(p^2-m^2)
    (p\cdot v-\Delta)} \nonumber\\
    &=&{\Delta\over16\pi^2}\left[\{
    (m^2 -{2\over3} \Delta^2)\frac1{\epshat} +
     ({4\over3} m^2 -\frac{10}{9}\Delta^2) + J_1(m,\Delta)\}g^{\mu\nu}
      \right. \nonumber\\
    &&+\mbox{} \biggl.
    \{(2m^2 - {8\over3}\Delta^2)\frac1{\epshat} +
     {4\over9}(7\Delta^2 - 3m^2) +
    J_2(m,\Delta)\}v^\mu v^\nu\biggr]\, .
\end{eqnarray}
The remaining integrals can be obtained by differentiation with
respect to $m^2$,
which will be denoted with a prime.  The definitions
of the functions $J_1$ and $J_2$ can be found in
Ref.~\citen{Glenn}.
For my purpose I need only the limiting values
\begin{eqnarray}
J(m, 0) &=&
   	{2\pi\over 3} m^3 \\
{\partial J(m, 0)\over \partial \Delta} &=&
	-m^2 \log(m^2/\mu^2),
\end{eqnarray}
where I have defined $J(m, \Delta) = \Delta J_1(m, \Delta)$.

\begin{figure}
\ifIncludeFigs
  \centerline{\epsffile{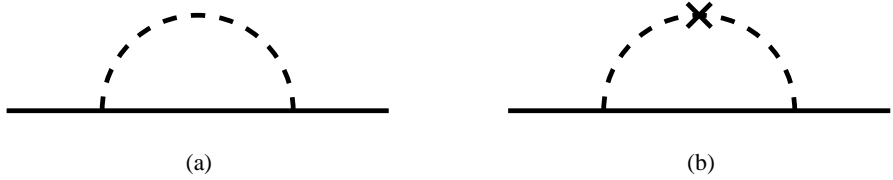}}
\fi
\caption{The diagrams which contribute to the heavy meson self energy.
Solid lines represent heavy mesons, dashed lines represent light mesons
and the cross represents an insertion of the ``hairpin'' vertex.
\label{fig:se}}
\end{figure}

The graphs which contribute to the self-energy are
shown in Fig.~\ref{fig:se}.
%
In diagram 1a, the ghost mesons will cancel the contribution from
the real mesons unless one of the vertices involves the singlet field.
Combining this with the contribution of the hairpin vertex diagram
1b,
I obtain
%
\eqb
\label{eqn:singletloop}
i\,\Sigma(v\cdot k) =  {6 i \over 16 \pi^2 f^2}\left[
  (2 g \gamma - {1\over3}\, g^2\alpha_0) J(M_d,  - v \cdot k )
 +  {1\over3}\, g^2 (m_0^2 - \alpha_0 M_d^2) J'(M_d, - v \cdot k )
 + \,\ldots\,
  \right].
\eqe
The terms not shown are analytic in $M_d$ and can be obtained
from Eq.~(\ref{eqn:Jint}) above.

\subsection{Wavefunction Renormalization and Decay Constants}

\begin{figure}
\ifIncludeFigs
  \centerline{\epsffile{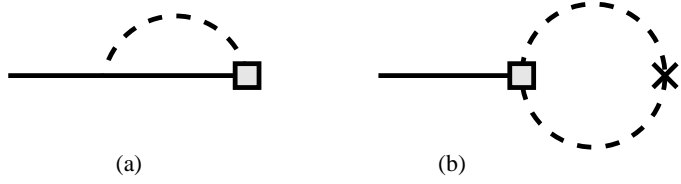}}
\fi
\caption{Corrections to the weak current vertex.  The
box represents an insertion of the weak current.
\label{fig:vertex}}
\end{figure}

The wavefunction renormalization constants are obtained by
differentiating the self-energy with respect to $2v\cdot k$
and evaluating on-shell.
I find
\begin{eqnarray}
  Z &=& 1 +  3 g^2 \Lq + 6 (g \gamma -
	{1\over3}\,g^2 \alpha_0)\LM + k_1 m_d.
\end{eqnarray}
Here and below, it is convenient to adopt the definitions
\begin{eqnarray}
\mu_d &=& \onpif{M_d^2} \log({M_d^2\over \mu^2}), \qquad
\mu_q \,=\,
\onpif{m_0^2/3}\log({M_d^2\over \mu^2}),
\nonumber\\
\mu_P &=& \onpif{m_P^2} \log({m_P^2\over \mu^2}),
 \qquad (P = \pi, K, \eta).
\end{eqnarray}
Loop corrections to the left-handed current vertex arise from the
diagrams of Fig.~\ref{fig:vertex}.
It is easy to see that the diagram
Fig.~\ref{fig:vertex}a vanishes: the loop integral must be proportional
to $v^\mu$, which will vanish when contracted with the projection
operator in the numerator of the $B^*$ propagator.
The remaining tadpole graph Fig.~\ref{fig:vertex}b
yields 
\eqb
{\im \alpha v^\mu \over 2 }
	\onpif{(-2\alpha_0 M_d^2 + m_0^2)/3} \log({M_d^2\over \mu^2}).
\eqe
The final results for the decay constants
are then found by combining
the wavefunction and vertex corrections:
\begin{equation}
\sqrt{M_{B}}f_{B} =
  \alpha \left[ 1 - {1\over2} (1 + 3g^2)\Lq
	- (3 g \gamma - (1+3g^2){\alpha_0\over3})\LM + \kappa_1 m_d \right].
\end{equation}
In contrast, the results in the full theory
are\cite{Grinsteinetal,Goity}
\begin{eqnarray}
\sqrt{M_{B_d}}f_{B_d} &=&
	\alpha \left[ 1 - {1\over2} (1 + 3g^2)\left(
		{3\over2} \L{\pi} + \L{K} + {1\over6}\L{\eta}\right)
		+\kappa_1 m_d + \kappa_2(m_s + 2m_d)
	        \right], \nonumber\\
\sqrt{M_{B_s}}f_{B_s} &=&
	\alpha \left[ 1 - {1\over2} (1 + 3g^2)\left(
		2\L{K} + {2\over3} \L{\eta} \right)
		+\kappa_1 m_s + \kappa_2(m_s + 2m_d)
		\right].
\end{eqnarray}

\subsection{Masses}
The correction to the mass is obtained by evaluating
the self-energy on-shell and removing the wavefunction renormalization
constant (though to the order I am working it does not contribute).
Defining
\eqb
M_B = \bar M_B - {3\over4} \Delta + \delta M_B
\eqe
where $\bar M_B$ is the spin-averaged mass in the chiral limit,
$\Delta$ is the hyperfine splitting and
$\delta M_B$ is the light-quark dependent contribution to the
mass,
I find
\eqb
 \delta M_{B^0} = 2\lambda_1\, m_d
 -  \onpif{2\pi}\left( g^2\, {m_0^2\over 3} M_d +
     (2g \gamma - 5g^2 {\alpha_0\over3}) M_d^3
     \right).
\eqe
while in the unquenched theory\cite{Goity,Jenkins}
\eqb
\delta M_{B_s} - \delta M_{B_d} =
  2\lambda_1 (m_s-m_d) -
  	\onpif{\pi g^2}(- 3m_\pi^3 + 2 m_K^3 + m_\eta^3).
\eqe

\subsection{Mixing}
\begin{figure}
\ifIncludeFigs
  \centerline{\epsffile{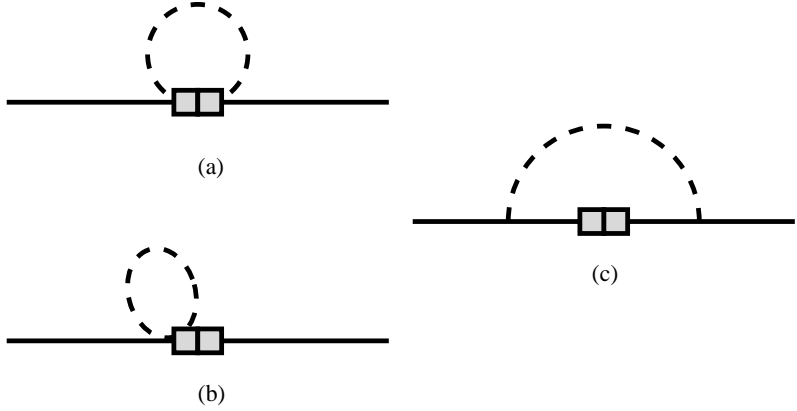}}
\fi
\caption{The corrections to $B-\bar B$ mixing.  The double box
represents an insertion of the mixing operator.  Corresponding
diagrams with hairpin vertices are not shown.
\label{fig:mixing}
}
\end{figure}

The constant $B_{B_a}$ is defined as the ratio
\eqb
B_{B_a} =
{ \bra{ \bar B_a}\bar q_L^a \gamma_\mu b_L
		\,\bar q_L^a \gamma^\mu b_L\ket{B_q}
  \over
  {8\over3} \bra{ \bar B_a}\bar q_L^a \gamma_\mu b_L\ket0
   \bra0 \bar q_L^a \gamma^\mu b_L\ket{B_a}} =
{ \bra{ \bar B_a}\bar q_L^a \gamma_\mu b_L
		\,\bar q_L^a \gamma^\mu b_L\ket{B_q}
  \over {2\over3} f_{B_a}^2 m_{B_a}^2\, B_{B_a}}.
\eqe
As shown by Grinstein and collaborators\cite{Grinsteinetal},
in the effective theory the operator
\eqb
\label{eqn:OB}
\bar q_L^a \gamma_\mu b_L\, \bar q_L^a \gamma^\mu b_L
\eqe
is represented by
\eqb
4\beta \trd\left[(\xi \bar H^{(b)} )^a \gamma_\mu L \right]
	\trd\left[(\xi H^{(\bar b)} )^a \gamma^\mu L \right],
\eqe
which is essentially just the square of the left-handed current.
The one-loop corrections to this operator
are shown in Fig.~(\ref{fig:mixing}).
There are two types of tadpoles which arise from the operator
Eq.~(\ref{eqn:OB}): those where each $\xi$ is expanded to $O(\phi)$
(Fig.~\ref{fig:mixing}a)
and those where only one of the $\xi$'s is expanded
(Fig.~\ref{fig:mixing}b).  The latter
tadpoles just renormalize $f_B$ and will cancel in the ratio for
$B_B$.
Thus it is only necessary to consider the former.
I find
\eqb
B_{B_d} = 4 \beta \left[1 - (1-3g^2)\left((1-{2\over3}\alpha_0) \LM
	+\Lq\right) + 	6 g \gamma \LM  + \beta_1 m_d \right],
\eqe
which should be compared with the unquenched
results\cite{Grinsteinetal}
($\beta_1$ and $\beta_2$ are additional counter-terms)
\begin{eqnarray}
B_B & = & 4 \beta \left[1 - (1-3g^2){2\over3}\L{\eta}
		+ \beta_1 m_d + \beta_2 (m_s + 2m_d)
	\right], \\
B_{B_s} &=& 4 \beta \left[1 -
    (1-3g^2)\left({1\over2}\L{\pi} + {1\over6}\L{\eta}\right)
		+ \beta_1 m_s + \beta_2 (m_s + 2m_d)
	\right].
\end{eqnarray}

The reader will note that in contrast to the earlier results,
$B_B$ has true chiral logarithms even in the absence of the
singlet coupling $\gamma$ and the kinetic coupling $\alpha_0$.
A similar phenomenon occurs in $B_K$,
as shown by Sharpe\cite{Sharpe:BK}.  The reason is that flavor
conservation allows disconnected quark loops to appear only
in the guise of the $\eta'$, so that even in full QCD they do
not contribute.

\subsection{Isgur-Wise Function}
The heavy quark current which mediates the decay $B\rightarrow D e\nu$
is represented at leading order by
\eqb
\xi_0(w) \trd[ \bar H_a^{(c)}(v')\gamma_\mu L H_a^{(b)}(v)],
\eqe
where $\xi_0(w)$ ($w = v \cdot v'$) is the leading-order
Isgur-Wise function.
In the full theory,
the leading corrections are\cite{Goity,JenkinsSavage}
(here the counter-terms $\eta_1$ and $\eta_2$ are functions of $w$)
\begin{eqnarray}
\xi_{u,d} &=& \xi_0(w) \left[ 1 +
	2g^2 (r(w)-1)\left(
		{3\over2} \L{\pi} + \L{K} + {1\over6} \L{\eta}
	\right) \right.\nonumber\\
  &&\phantom{\xi_0(w)}+ \left.
	2(r(w)-1)\biggl(\eta_1(w) m_d + \eta_2(w)(m_s+2m_d)\biggr)
	\right],
  \nonumber\\
\xi_s &=& \xi_0(w) \left[1
	+ 2g^2(r(w) - 1)\left(
		2\L{K} + {2\over3} \L{\eta}\right) \right.\nonumber\\
  &&\phantom{\xi_0(w)}+
  	\left. 2(r(w)-1)\biggl(\eta_1(w) m_s + \eta_2(w)(m_s+2m_d)\biggr)
	\right],
\end{eqnarray}
where
\eqb
r(w) = {1\over \sqrt{w^2-1}}\log(w^2 + \sqrt{w^2-1}).
\eqe
The quenched results take the by-now-expected form
\eqb
\xi_d(w) = \xi_0(w)\left[1 + 2(r(w)-1)\left( g^2\Lq +
	(2g \gamma - {2\over3} g^2\alpha_0)\LM
	+ \eta_1(w) m_d \right)
	\right].
\eqe

\section{Discussion and Numeric Results}

To obtain numeric values it is necessary to know the values
of the various couplings which enter the Lagrangian.
%
Combining data on the $D^*$ width and branching
fractions\cite{ACCMOR:Dstar,CLEO:Dstar},
Amundson \etal\ Ref.~(\citen{Amundson:g}) obtained the constraint
$0.1 < g^2 < 0.5$.  The spread is caused by the uncertainty
in the branching fraction $BR(D^{*+}\rightarrow D^+\gamma)$;
taking the central value yields $g\simeq 0.5$.
QCD sum rules\cite{QCDSR:g1}
and relativistic quark models\cite{RQM:g1}
favor a smaller value, $g \sim 1/3$.
Given this uncertainty,
I will show results for different values of the coupling.
There is no information on the coupling $\gamma$,
but $1/N_c$ arguments suggest that it is small.%
\footnote{
  The same argument implies a
  suppression of the singlet coupling to the nucleon.  This
  is confirmed in a phenomenological study by Hatsuda\cite{Hatsuda},
  who found $g_{\eta' NN} \lesssim 1.1$, which should be compared
  with $g_{\pi NN} = 13.4$.
}
They also suggest that $\alpha_0$ is
small; direct evidence from $\eta-\eta'$ mixing confirms this.
Consequently, I will take both $\gamma$ and $\alpha_0$ to
vanish.\footnote{
	Even if $\gamma$ is as large as $g/3$, I find it
	changes the quenched results by only $5\%$ or so.
}
The maximum value of $w$ in the decay $B\goesto D l \nu$
is about $1.8$, so I will
use $r(1.8) = 0.76$ when evaluating $\xi$.
Finally, I will choose $\mu = 1\GeV$.

There are several ways to determine $m_0$, each
giving a different result.
The Witten-Veneziano large $N_c$ formula\cite{WV}
$m_0^2 = m_{\eta'}^2 + m_\eta^2 - 2 m_K^2$
gives $m_0 \approx 852\MeV$,
while from the $\eta-\eta'$ mass splitting
Sharpe\cite{Sharpe:BK} estimated $m_0 \approx 900 \MeV$.  It has
also been computed directly on the lattice.
Early attempts\cite{M0old} found
$m_0 \approx 570-920\MeV$, but with limited statistics and
a strong dependence on the lattice spacing.
Recently, a more accurate computation has been performed.
Kuramashi {\it et al.}\cite{M0new} extracted $m_0$
by comparing the one and two loop contributions to the $\eta'$
propagator;  they found $m_0 = 751(39)\MeV$.  Using the $U(1)$ Ward
identity relation $m_0^2 = 6 \chi/f_\pi^2$, with $\chi$ the
topological susceptibility, the same group found $m_0 = 1146(67)$ with
$\chi$ and $f_\pi$ obtained on the same lattice.  They attributed
this larger result to contamination
from extra terms in the Ward identity
induced by the use of Wilson fermions.
I will choose $m_0 = 750\MeV$, but will also show some results
for $m_0 = 1100\MeV$.

For an honest calculation, it is also necessary to specify
the $O(m_q)$ counter-terms.
But it is an unfortunate
fact that there is little to constrain them, save the general
expectation that their natural scale is $\Lambda_\chi \approx 4 \pi f$.
A common practice when confronted with this situation to
assume that the counter-terms are overshadowed by the logarithmic
contributions.
Another approach is to reduce
the dependence on these unknown terms by taking appropriate
ratios.
Fortunately,
since it is expected that the coefficients in
the chiral expansion should be (almost) the same in the
quenched
and full theories, some of the counter-terms will cancel
when the predictions of the two theories are subtracted
to compute the error.
In particular, the errors
in $f_{B_s}/f_{B_d}$, $\xi_s/\xi_d$, $B_{B_s}/B_{B_d}$
and $\delta M_{B_s} - \delta M_{B_d}$ are
free of counter-terms.
Results for these quantities
are shown in Table \ref{tab:ratios};
in order to illustrate the $m_0$ dependence,
the quenched results are shown at both $m_0 = 750\MeV$
and $m_0 = 1100\MeV$.
\def\arraystretch{1.3}
\begin{table}[tb]
\caption{\label{tab:ratios}
	Quenched and full heavy-light quantities.
	Only the non-analytic part of the mass difference is shown.
}
\begin{center}
\begin{tabular}{|lrrr|}
\hline
Quantity          &    Unquenched       &    $m_0 = 750\MeV$ &
$m_0 = 1100\MeV$ \\
\hline
$f_{B_s}/f_{B_d}-1$ & $ 0.074(1+3g^2)$ & $-0.11(1+3g^2)$ &
	$-0.23(1+3g^2)$ \\
\hline
$B_{B_s}/B_{B_d}-1$ & $ 0.052(1-3g^2)$ & $-0.11(1-3g^2)$ &
	$ -0.36(1-3g^2)$ \\
\hline
$\xi_s/\xi_d -1$     & $ 0.059\,g^2 $    & $ - 0.086\,g^2 $ &
	$-0.18\,g^2 $ \\
\hline
$\delta M_{B_s} - \delta M_{B_d}$ &
	$ -450\, g^2 \MeV$	& $ -340\, g^2 \MeV$
		& $ -740\, g^2 \MeV$  \\
\hline
\end{tabular}
\end{center}
\end{table}
The ratios in the
quenched theory are computed by substituting
$M_s = \sqrt{2m_K^2 - m_\pi^2} = 680\MeV$ and $M_d = m_\pi$.
Concentrating on the results at $m_0 = 750\MeV$, one sees that
the corrections to the ratios
are similar in magnitude but opposite in
sign to those in the full theory.  This
is a result of the fact that the quenched logarithms
diverge in the chiral limit.  Notice that the
corrections for the mass
splittings threaten to be larger than the splittings themselves
unless $g$ is small.  This suggests either a large cancelation
occurs with the leading $\lambda_1$ term or higher-order
corrections are important.  Either solution casts
doubt on the reliability of the error estimates in this case.

\begin{figure}[tb]
\ifIncludeFigs
  \epsfxsize=.9\columnwidth
  \centerline{\epsffile{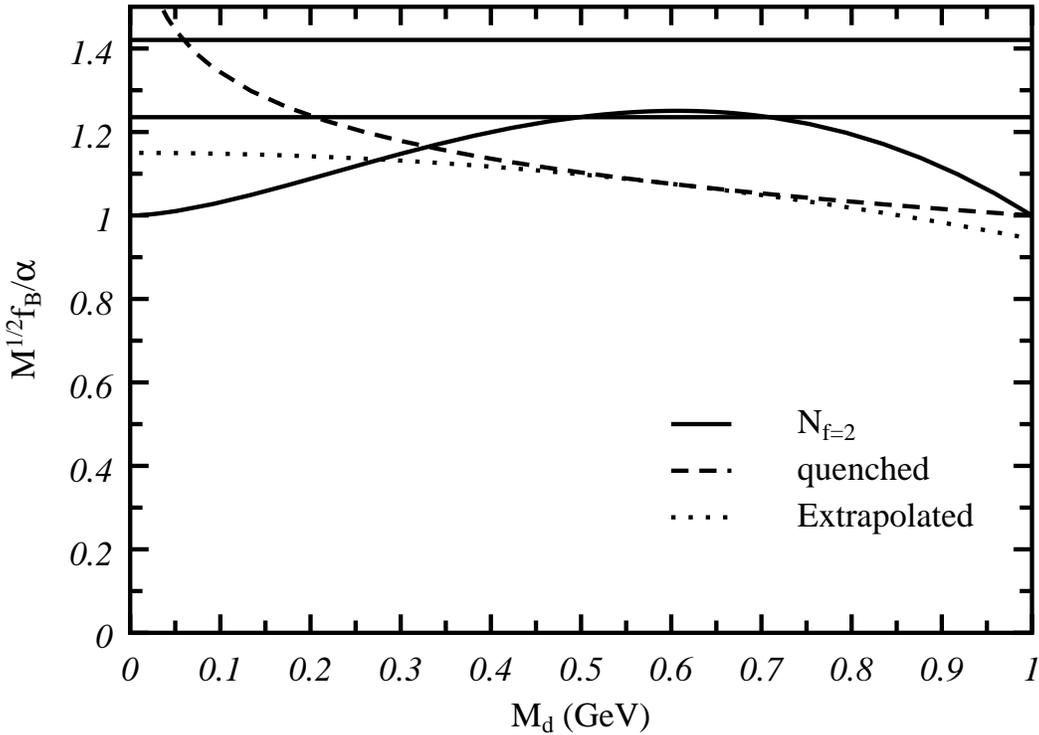}}
\fi
\caption{\label{fig:fbvsm}
	Quenched and unquenched corrections to $f_B$ as a function of
	meson mass.  Also shown is a linear (in the quark, not the
	meson mass) extrapolation of the quenched $f_B$.  The horizontal
	lines are the predictions for $f_{B_s}$ and $f_B$ in the
	full theory.}
\end{figure}

While the results in Table \ref{tab:ratios} suggest large
quenching errors --- particularly in $f_{B_s}/f_B$ --- it is likely that
the error in actual simulations will be less.
The reason is the following.
Currently, most simulations are performed with
quark masses corresponding to pion masses in the range
$400 \lesssim M_\pi \lesssim 1000 \MeV$.  The results are then
extrapolated linearly (in the quark mass) to the chiral limit
and the physical $m_\pi$.
Due to the familiar property of the logarithm,
quenched loop corrections change as much in the interval
$140 < M_d < 350 \MeV$ as they do in the interval $350 < M_d < 1000\MeV$.
Consequently, in the mass range covered by lattice simulations
the quenched logarithm appears linear.  This can
be seen in Fig.~\ref{fig:fbvsm}, where both the
``true'' quenched and linearly extrapolated predictions
for $f_B$ are shown.
%
Clearly, the two cannot be
distinguished for masses greater than say $300 \MeV$, but
the extrapolated result underestimates the ``true'' behavior
by more than $10\%$ at $M_d = m_\pi$.
While the primary motivation for this extrapolation is the desire
to efficiently invert the quark propagators,
it has the side-effect of reducing the quenching errors.
Moreover, it is the correct thing to do, since
the goal is to describe unquenched QCD and quenched ChPT
clearly fails to do this in the chiral limit.
Table \ref{tab:error1} compares the two methods of computing
the error at different values of the coupling.
\begin{table}[tb]
\caption{\label{tab:error1}
	The errors for
	different choices of the coupling $g$.  Results are shown
	using both the exact and extrapolated quenched predictions.
 	An error is negative when the quenched quantity is smaller than
	the full.
}
\begin{center}
\begin{tabular}{|lrrrrrr|}
\hline
Errors& \multicolumn{3}{c}{Exact} & \multicolumn{3}{c|}{Extrapolated} \\
		& $g=0.7$ & 0.5 & 0.33 & 0.7 & 0.5 & 0.33\\
\hline
$f_{B_s}/f_{B_d}$
	  & $-0.45$    & $-0.32$       & $-0.24$
	  & $-0.28$    & $-0.20$ & $-0.15$  \\
\hline
$B_{B_s}/B_{B_d}$
	& $0.078$        & $-0.041$      & $-0.11$
	& $0.012$      & $-0.0062$      & $-0.017$ \\
\hline
$\xi_s/\xi_d$
	& $-0.072$      & $-0.036$       & $-0.016$
	& $-0.045$     & $-0.023$      & $-0.010$ \\
\hline
$M_{B_s}-M_{B_d}$
	& $49 \MeV$   & $25 {\rm \,MeV}$   & $11 {\rm \,MeV}$
	& -  & - & - \\
\hline
\end{tabular}
\end{center}
\end{table}
One sees that the error is substantially
reduced by the extrapolation.
The errors in $B_B$ and $\xi$ are
relatively small to start with and become negligible when extrapolated.
However, even with the extrapolation the error in $f_B$ is
larger then one might have hoped.  Note also that $f_{B_s}/f_B$
is {\em smaller} in the quenched theory.

In fact, the size of the error in $f_B$ is easy to understand.  For
the Isgur-Wise function, the error is small because the corrections
themselves are small.  Conversely, the corrections to $f_B$ are
large and so the error is large.  Moreover, they
are driven by the tadpole terms, which remain large even if the
coupling $g$ vanishes.  The tadpoles, however, do not depend on
the heavy quark mass, so it should be possible to eliminate
them by studying the
$1/M$ corrections\cite{Booth2}.

Some additional understanding of the
differences between the quenched and
dynamical theories
may be gained by
comparing them in the mass range probed on the lattice.
For this it may be better to consider
a two quark theory with degenerate masses (rather than the
full three-flavor theory), since it is closer to
the type of theory studied in unquenched simulations.
In Fig.~\ref{fig:fbvsm} the predictions for $f_B$ are shown.
Here I have neglected the unknown counter-terms, though it
is clear from the graph
that there must be a positive term of $O(m_d)$ since
lattice simulations find that $f_B$ increases with
the light quark mass.
There are two general features that should be noted in
Fig.~\ref{fig:fbvsm}.  First, that in both the quenched
and two-flavor theories, $f_B$ is less than both $f_{B_d}$
and $f_{B_s}$.  This may be attributed to the fact that the full
theory has more mesons contributing to loop corrections.  The
second observation is that the gap between the two-flavor and
quenched corrections grows as $M_d$ increases toward the point
$M_s = \sqrt{2m_K^2-m_\pi^2}$.  Thus, the fact that quenching
decreases the ratio $f_{B_s}/f_{B_d}$ is due to the different
nature of the quenched logarithm.
Finally, Table~\ref{tab:latmass} compares the quenched
and two-flavor predictions at a representative mass
of $M_d = 600\MeV$
(again neglecting counter-terms).
\begin{table}[tbh]
\caption{\label{tab:latmass}
	Quenched and unquenched results at
	$M_d\, (= m_\pi) = 600\MeV$.
}
\begin{center}
\begin{tabular}{|lrr|}
\hline
Quantity & $m_0 = 750\MeV$
	& Unquenched ($N_f=2$) \\
\hline
$\sqrt{M} f_B/\alpha -1 $ &
   $ 0.031(1+3g^2)$ & $ 0.10(1+3g^2)$ \\
\hline
$B_B/4\beta - 1$ &
   $ 0.20(1-3g^2)$  & $ 0.069(1-3g^2)$ \\
\hline
$\xi_d/\xi_0 - 1$ &
   $ 0.025\,g^2 $   & $ 0.083\,g^2 $ \\
\hline
\end{tabular}
\end{center}
\end{table}
It can be seen that the errors are comparable to those found in the
extrapolated ratios.

\section{Conclusions}

I have included heavy mesons into the framework of quenched chiral
perturbation theory and used it to study the error arising from
the use of the quenched approximation in lattice studies
of heavy-light mesons.  These lattice studies
are important for the phenomenology of the standard model.
Because estimates of the error
depend on the as-yet unmeasured value of the the $B^*B\pi$
coupling $g$, results were shown for several
values of $g$ in the allowed range.  It was seen that the
errors in $B_B$ and $\xi(w)$ were negligible.
However, the error in $f_B$ was surprisingly
large, more than $15\%$ in the best case.  It was observed that
the large error follows from the large corrections present
in both theories.  These large corrections were traced to the
tadpole corrections, and it was suggested that they might
be eliminated by studying the $1/M$ corrections to the theory.
Indeed, Grinstein\cite{GrinsteinR,Glenn}
has advocated doing just that in the continuum by studying
the double ratio ${ f_{B_s}/f_{B} \over f_{D_s}/f_D}$.
This is done for
the quenched theory in a forthcoming work\cite{Booth2}.


A general conclusion that can be drawn from the quenched chiral
calculation is that quenching tends to decrease the ratio $f_{D_s}/f_D$.
This is in agreement with the one
unquenched simulation\cite{HEMCGC:fB}, which found $f_{D_s}/f_D = 1.34$,
a value larger than that typically found in quenched calculations.
This may have implications for reconciling lattice predictions of
$f_{D_s}$ with the recent CLEO measurement\cite{CLEO:fDs}.

In the future it would be interesting to study heavy baryons
containing two light quarks within this framework.  Because of the
presence of two light quarks, the quenched theory will be less trivial
and the loop corrections will have a more complicated flavor
structure, more closely resembling that of the light mesons.
It would also be useful to to move beyond $1/N_c$ arguments
for the magnitude of $\gamma$.  It appears that it could be
calculated within QCD sum rules using the same techniques
that have recently been applied to $g$\cite{QCDSR:g1}.

\acknowledgements
I would like to think the University of Chicago theory group for
its hospitality while this work was completed.
This work was supported in part by DOE grant DE-FG05-86ER-40272.

\appendix
\section{Renormalization Constants}

Within the context of dimensional regularization,
the singularities of the effective Lagrangian are customarily
described in terms of the parameter $L(\mu)$ which contains
the singularity at $D=4$
(recall $D = 4 - 2\epsilon$,
  $1/\epshat =  1/\epsilon + \log 4\pi - \gamma_{\rm E} + 1$):
\eqb
L(\mu) = {1\over 16\pi^2} \mu^{-2\epsilon}{1\over \hat \epsilon}.
\eqe
It is convenient to decompose an arbitrary
coupling $k$ as $k = k^{\rm r}(\mu) + \bar k\, L(\mu)$.

To render the Lagrangian Eq.~(\ref{eqn:FullL}) finite, it is
necessary to add the counter-term
\eqb
3\,g^2 {m_0^2 \over f^2}
    \trd\left[\bar H_a i v\cdot D_{ba} H_b\right]\,L(\mu). 
\eqe
In addition, $\bar k_1$ must be taken to be
\eqb
\bar k_1 =
  6 (g \gamma - {1\over3}\,g^2 \alpha_0) {2\mu_0\over f^2}.
\eqe
The current Eq.~(\ref{eqn:currentfull})
is renormalized with
\eqb
\bar \kappa_1 =  -
	((1+3g^2)\frac{\alpha_0}3 - 3 g\gamma)\,{2\mu_0\over f^2}.
\eqe
and in addition $\alpha$ must be rescaled:
\eqb
\alpha^{\rm r}(\mu) = \alpha\left[ 1 +
	{1\over2}(1+3g^2){m_0^2/3\over f }\,L(\mu)\right].
\eqe
The description of $B-\bar B$ mixing requires the counter-term
(no sum on $a$)
\eqb
4\,\beta\, \beta_1 \trd\left[(\xi \bar H^{(b)} )^a \gamma_\mu L \right]
  \trd\left[(\xi H^{(\bar b)} )^a \gamma^\mu L \right] (\scr{M}_+)_{aa},
\eqe
and the couplings must be
\begin{eqnarray}
\beta^{\rm r}(\mu)&=&
	\beta\left(1+ (1-3g^2) {m_0^2/3 \over f} L(\mu) \right), \\
\bar \beta_1 &=& \left( (1-3g^2)(1-{2\over3}\alpha_0) - 6g\gamma\right)
	{2 \mu_0 \over f^2}.
\end{eqnarray}
Finally, the Lagrangian for $b \goesto c$ transitions needs the
counter-term
\eqb
\eta(w) (r(w)-1)
   \trd[ \bar H_a^{(c)}(v')\gamma_\mu L H_b^{(b)}(v)] (\scr{M}_+)_{ba},
\eqe
with
\eqb
\bar \eta(w) =  - 4(g\gamma-{1\over3} g^2\alpha_0)
	\,{2 \mu_0 \over f^2}
\eqe
and the rescaled coupling
\eqb
\xi_0^{\rm r}(w,\mu) = \xi_0(w)
  \left(1 + 2(r(w)-1)\,g^2 {m_0^2/3 \over 3} L(\mu) \right). \\
\eqe

\def\jvp#1#2#3#4{#1~{\bf #2}, #3 (#4)}
\def\PR#1#2#3{\jvp{Phys.~Rev.}{#1}{#2}{#3}}
\def\PRD#1#2#3{\jvp{Phys.~Rev.~D}{#1}{#2}{#3}}
\def\PRL#1#2#3{\jvp{Phys.~Rev.~Lett.}{#1}{#2}{#3}}
\def\PLB#1#2#3{\jvp{Phys.~Lett.~B}{#1}{#2}{#3}}
\def\NPB#1#2#3{\jvp{Nucl.~Phys.~B}{#1}{#2}{#3}}
\def\SJNP#1#2#3{\jvp{Sov.~J.~Nucl.~Phys.}{#1}{#2}{#3}}
\def\AP#1#2#3{\jvp{Ann.~Phys.}{#1}{#2}{#3}}
\def\PL#1#2#3{\jvp{Phys.~Lett.}{#1}{#2}{#3}}
\def\NuovoC#1#2#3{\jvp{Nuovo.~Cim.}{#1}{#2}{#3}}
\def\NPBPS#1#2#3{\jvp{Nucl.~Phys.~B~(Proc.~Suppl.)}{#1}{#2}{#3}}
\def\Prog#1#2#3{\jvp{Prog.~Theor.~Phys.}{#1}{#2}{#3}}
\def\ZPC#1#2#3{\jvp{Z.~Phys.~C}{#1}{#2}{#3}}




\end{document}